\documentclass[12pt]{iopart}
\usepackage{graphicx}
\usepackage{natbib}
\def \half{\textstyle{1 \over 2}}
\begin{document}

\title[A quantisation of time]{A quantisation of time}

\author{J C Jackson}
\address{16 The Park, Newark, Nottinghamshire, UK\newline\newline Received 5 July 1977}

\begin{abstract}
A model of discrete space-time is presented which is, in a sense, both Lorentz 
invariant and has no restriction on the relative velocity between particles
(except $v<c$). The space-time has an inbuilt indeterminacy.
\end{abstract}

\section{Introduction}

Through out the development of man's ideas about the physical world, the view that 
time and space are continuous has generally prevailed. Discrete models have occasionally 
been entertained, with space-time events labelled by integral coordinates 
(Schild 1948), but these have had virtually no impact on physics. The continuous 
picture certainly provides an accurate description of affairs on the macroscopic scale, 
but there are strong advocates (notably Chew 1963; Penrose 1967) of the view that this
picture must become inaccurate at the sub-microscopic level, probably when we are
dealing with intervals shorter than those encountered in elementary particle physics. 

As a simple model of discrete space-time, Schild (1948) has considered a hyper-
cubic lattice, comprising all events in Minkowski space-time whose four coordinates 
$(t,x,y,z)$ are integers. The problem is that this is not a Lorentz invariant picture,
as a general Lorentz transformation destroys the hypercubicity. Schild preserves Lorentz 
invariance by allowing only those transformations which preserve the integer labelling. 
Unfortunately the smallest non-zero velocity allowed by the resulting group is 
$3^{1/2}c/2$ where $c$ is the velocity of light, which is not very promising as far as
physical applications are concerned. The purpose of this paper is to describe a discrete
model which, in a sense to be defined, is both Lorentz invariant and allows any relative 
velocity between $0$ and $c$.
\vspace{2.1cm}

\noindent
\leftline{E-mail: john.jackson@northumbria.ac.uk; jcj43@talktalk.net}
\newpage

\section{The $k$-calculus}

\begin{figure}[here]
\begin{center}
\includegraphics[scale=0.75]{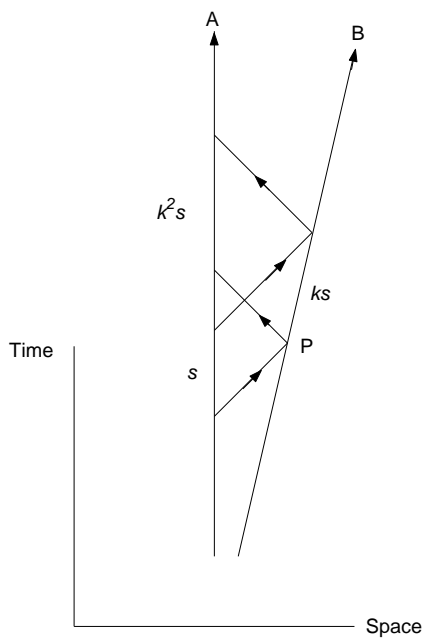}
\caption{The $k$-calculus.}
\label{TimeFig1}
\end{center}
\end{figure}

Bondi (1965) has presented a formulation of special relativity based upon what he 
calls the $k$-calculus, which provides a very convenient framework for my ideas.
He considers observers equipped only with clocks and light sources; these observers 
assume a fixed value for the velocity of light, and measure distances using a radar 
technique, by observing the interval between emission of light pulses and reception of 
the corresponding echoes. Naturally the space-time so constructed is Minkowski 
space-time; the point of the exercise is that many of the elementary results of special 
relativity follow in an almost trivial manner, without having to derive the Lorentz 
transformation first. For example, Figure \ref{TimeFig1} shows two such observers A and B moving 
with relative velocity $v$. To measure this velocity, A sends out two flashes of light 
separated by an interval $s$ on his clock. On B's clock, the corresponding interval
between reception of the two flashes is $ks$, which is Bondi's definition of the $k$-factor; 
by symmetry A  must observe an interval of $k^2s$ between reception of the echoes.
If A emits the first pulse when his clock shows time $S_\mathrm{e}$, and receives the first
echo at time $S_\mathrm{r}$, then he assigns coordinates $(T,X)$ to event P given by
(using units in which $c=1$)

\[
T=\half(S_\mathrm{r}+S_\mathrm{e})
\]
\vspace{-0.45cm}
\[
X=\half(S_\mathrm{r}-S_\mathrm{e}).
\]

\vspace{-0.6cm}

\vspace{0.75cm}
\noindent
Denoting changes by $\Delta T$, $\Delta X$, etc, we have
\newpage

\[
\Delta T=\half(\Delta S_\mathrm{r}+\Delta S_\mathrm{e})=\half(k^2+1)s
\]
\vspace{-0.45cm}
\rightline{(1)}
\vspace{-0.45cm}
\[
\Delta X=\half(\Delta S_\mathrm{r}-\Delta S_\mathrm{e})=\half(k^2-1)s
\]

\noindent
giving the velocity
\[
v=\Delta X/\Delta T=(k^2-1)/(k^2+1)
\]

\noindent
or
\[
k=\left({1+v \over 1-v}\right)^{1/2}.
\]

\vspace{0.5cm}
\noindent
Hence in this simple fashion the relativistic Doppler shift formula has been derived.

\section{Discrete time}

The space-time to be described here is constructed in exactly the above fashion, but 
assumes that proper-time intervals are discrete rather than continuous. Thus for 
example the interval between A's emission of the two flashes in the above experiment 
must be measured by a positive integer $n_\mathrm{e}$ rather than by a continuous real
number $s$. Similarly the interval between A's reception of the two echoes is measured
by another positive integer $n_\mathrm{r}$. We now define the $k$-factor characterising
B's motion by 

\[
k^2=n_\mathrm{r}/n_\mathrm{e}
\]
\vspace{-0.9cm}
\rightline{(2)} 

\vspace{0.3cm}
\noindent
and B's velocity relative to A by 

\[
v=(n_\mathrm{r}-n_\mathrm{e})/(n_\mathrm{r}+n_\mathrm{e}).
\]
\vspace{-0.9cm}
\rightline{(3)} 

\vspace{0.4cm}
\noindent
At this stage, the model runs into severe difficulties as a model of the real world,
as equations (2) and (3) restrict $k^2$ and $v$ to rational values. Moreover, the
result of a  determination of B's velocity will depend upon which pulse emission
interval A decides to use. For example, A might use a large value, say $n_\mathrm{e}=10^6$,
in one determination, and observe say $n_\mathrm{r}=10^6+2$, giving $v=10^{-6}$,
i.e.~$300$ m s$^{-1}$. In a second  determination A might choose a small interval,
say the extreme case $n_\mathrm{e}=1$; the smallest non-zero velocity in this case
is given by $n_\mathrm{e}=2$, and is $v=1/3$, i.e.~$10^8$ m s$^{-1}$.
 
These problems are similar to those encountered in Schild's model; they disappear if 
we drop the assumption that $n_\mathrm{r}$ is precisely determined by $n_\mathrm{e}$.
For example, observer A might conduct a series of experiments to determine the velocity
of a given particle B. In each experiment, A allows the same integer interval $n_\mathrm{e}$
between sending out the two light flashes. In the model advocated here, the values of
$n_\mathrm{r}$ observed in such a series need not be identical; this is taken as one
of the axioms of the model. Instead all values of $n_\mathrm{r}$ can be observed,
but do so with a frequency distribution which gives the correct mean value for the
ratio in equation (2). In the above particular example with $n_\mathrm{e}=1$,
the value $n_\mathrm{r}=1$ would be observed in almost all experiments, and 
$n_\mathrm{e}=2$ in only about three experiments per million conducted.
 
Thus we have a model of space-time with an inbuilt indeterminacy. All that is 
needed to complete the picture is a specification of the frequency distribution of the 
observed number $n_\mathrm{r}$. The natural choice is the Poisson distribution, which finds a 
number of applications in physics. For example, in a gas the number of atomic 
collisions experienced by any one atom in unit time is a random integer distributed in 
this fashion; if $P_{n_\mathrm{c}}$is the probability of making $n_\mathrm{c}$ such collisions,
we have
 
\vspace{0.3cm}
\[
P_{n_\mathrm{c}}=n_\mathrm{m}^{n_\mathrm{c}}\e^{-n_\mathrm{m}}/n_\mathrm{c}! 
\]

\vspace{0.3cm}
\noindent
where $n_\mathrm{m}$ is the mean number of collisions per unit time. I shall assume such a 
distribution for $n_\mathrm{r}$. In the series of experiments described above, we require

\vspace{0.3cm}
\noindent
\[
P_{n_\mathrm{e},n_\mathrm{r}}=(k^2n_\mathrm{e})^{n_\mathrm{r}}\e^{-k^2n_\mathrm{e}}/n_\mathrm{r}!
\]
\vspace{-0.9cm}
\rightline{(4)} 
  
\vspace{0.6cm}
\noindent
This is the probability that if A chooses a pulse emission interval $n_\mathrm{e}$,
he will observe an echo reception interval $n_\mathrm{r}$. The factor $k^2$ characterising
B's motion relative to A is now the mean value of the ratio $n_\mathrm{r}/n_\mathrm{e}$,
rather than the value resulting from a particular observation; $k^2$ is thus not restricted
to rational values. Similarly, the mean of the velocity defined by equation (3) can have any
value between $-1$ and $+1$. The standard deviation of $n_\mathrm{r}$ is exactly
$kn_\mathrm{e}^{1/2}$, so that $n_\mathrm{r}/n_\mathrm{e}\rightarrow k^2$ as
$n_\mathrm{e}\rightarrow \infty$; hence we recover the classical picture over sufficiently
long intervals, when the indeterminacy is of negligible proportions.

\section{The nature of time}

It may be that the above analogy with the kinetic theory of gases is more than just an 
analogy. I have a picture of space-time as in some sense comprising a sea of particles 
which I shall call chronons. Other particles float in this sea and sense a continual 
chronon bombardment; this sensation is called time. I have found this physical picture 
a great conceptual aid, but it is not essential to the model, the essence of which is 
contained in the following purely mathematical postulates: 

(i) Discreteness. Proper-time intervals are discrete and the structure of space-
time is given by the radar map.
 
(ii) A correspondence principle. The integers, measuring any two time intervals 
which classically are causally related in a linear manner, are randomly related 
in such a way that specifying the one only fixes the mean of the other, this 
mean to coincide with the classical value.
 
(iii) The distribution. The probability function specifying the random behaviour is 
the Poisson distribution.

\vspace{0.3cm}
\noindent
In what follows, I shall frequently use the expression `counting chronons'; it can be 
interpreted as meaning just the registering of discrete time elements; however, it may 
have a more literal meaning.

\section{Lorentz invariance}

\begin{figure}[here]
\begin{center}
\includegraphics[scale=0.75]{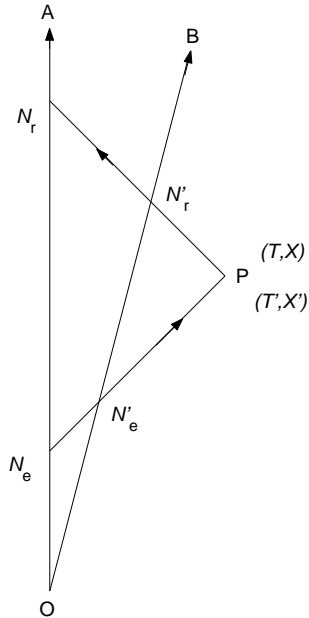}
\caption{The Lorentz Transformation.}
\label{TimeFig2}
\end{center}
\end{figure}

The postulates of this space-time model do not assign a special position to any 
particular class of observers, so that the model must basically be Lorentz invariant. 
However, the indeterminacy will mask the invariance in any one observation. It is instructive 
to examine the $k$-calculus derivation (Bondi 1965) of the Lorentz transformation in the light 
of the ideas expounded above. \thinspace Observers A and B in Figure \ref{TimeFig2} move with relative 

\vspace{0.5cm}
\noindent
motion characterised by a factor $k$, and agree to start their chronon 
counters as they pass each other at event O. After counting $N_\mathrm{e}$ chronons A emits a 
light pulse to illuminate event P, and receives the echo after counting $N_\mathrm{r}$ chronons. 
Observer B similarly illuminates event P, the corresponding chronon counts being $N_\mathrm{e}'$ 
and $N_\mathrm{r}'$. A and B respectively assign (half-integral) coordinates $(T,X)$ and
$(T',X')$ to this event, defined by

\[
T=\half(N_\mathrm{r}+N_\mathrm{e})~~~~~~X=\half(N_\mathrm{r}-N_\mathrm{e})
\]
\vspace{-0.45cm}
\[
T'=\half(N_\mathrm{r}'+N_\mathrm{e}')~~~~~~X'=\half(N_\mathrm{r}'-N_\mathrm{e}').
\]

\noindent
Writing
\[
N_\mathrm{e}'/N_\mathrm{e}=k_1~~~~~~N_\mathrm{r}/N_\mathrm{r}'=k_2
\]

\noindent
the equations
\[
T'=\half(k_1+1/k_2)T-\half(k_1-1/k_2)X
\]
\vspace{-0.9cm}
\rightline{(5)}
\vspace{0.15cm}
\[
X'=\half(k_1+1/k_2)X-\half(k_1-1/k_2)T
\]
\vspace{-0.9cm}
\rightline{(6)}

\noindent
and
\[
T'^2-X'^2=(k_1/k_2)(T^2-X^2)
\]

\noindent
are easily derived. Classically we would have

\[
k_1=k_2=k=\left({1+v \over 1-v}\right)^{1/2}
\]

\noindent
when equations (5) and (6) are just the Lorentz transformation. In general we have 
$k_1\not =k_2$, so that these equations do not coincide with a Lorentz transformation. 
However,the correspondence principle ensures that the mean values of $k_1$ and $k_2$, 
taken over many similar observations, are equal to $k$, so that in this sense Lorentz 
invariance is maintained. Of course in the case of the macroscopic intervals $k_1$
and $k_2$ are so nearly equal that the lack of invariance would not be noticed.

\section{Uncertainty}

Modern classical and quantum physics are troubled by infinities in a number of areas; 
the possibility that these divergences might disappear if a difierent space-time model 
were to be adopted, with the equations of classical and quantum physics suitably 
rewritten, has been one of the motivations behind the search for such alternatives. 
Given such a model, the appropriate reformulation of the laws might not be unambiguously
indicated. In this context, the present model offers almost an embarrassment 
of riches, as any reformulation of classical physics necessarily leads to laws which 
reflect the indeterminacy in the model. A reformulation of quantum theory, with its 
inherent uncertainty, within the present framework, might lead to too much indeterminacy.
A tempting point of view is that this model plus classical physics might be an 
alternative to classical space-time plus quantum theory. An intermediate viewpoint 
is that this model might make the `second quantisation' programme unnecessary. 
The present model is hardly sufficiently well developed to permit a general reformulation
of the kind discussed above. However, we can examine the most elementary 
classical dynamical concept, namely the one-dimensional motion of a free particle. 
Consider first a non-relativistic case: an observer A and an elementary particle B of 
mass $m$ at rest relative to A, in the sense that $k=1$ in the distribution (4).
A carries out a radar determination of B's velocity using a flash emission interval
$n_\mathrm{e}$; in general A will observe an echo reception interval
$n_\mathrm{r}\not =n_\mathrm{e}$, so that according to equation (3) B does not appear
to be at rest at all; over a series of such experiments, B has a mean 
square velocity $\langle v^2\rangle$ given by

\vspace{0.3cm}
\[
\langle v^2\rangle
=\langle\left({n_\mathrm{r}-n_\mathrm{e} \over n_\mathrm{r}+n_\mathrm{e}}\right)^2\rangle
\sim {{\langle (n_\mathrm{r}-n_\mathrm{e})^2 \rangle} \over 4n_\mathrm{e}^2}.
\]
\vspace{0.15cm}

\noindent
Distribution (4) with $k=1$ gives $\langle (n_\mathrm{r}-n_\mathrm{e})^2 \rangle=n_\mathrm{e}$,
giving

\vspace{0.15cm}
\[
\langle v^2\rangle=1/4n_\mathrm{e}
\]
\vspace{0.15cm}

\noindent
which constitutes an estimate of B's energy. As the true energy of B is zero, this
estimate is in error by an amount

\vspace{0.15cm}
\[
\Delta E=\half m\langle v^2\rangle\sim m/8n_\mathrm{e}.
\]
\vspace{0.15cm}

\noindent
This estimate refers to B's motion during a time interval for the observer of extent
$\Delta t=n_\mathrm{r}\sim n_\mathrm{e}$.  The product $\Delta E\Delta t$ is thus
given by

\[
\Delta E\Delta t\sim m/8.
\]
\vspace{-0.9cm}
\rightline{(7)}

\vspace{0.3cm}  
\noindent
If $n_\mathrm{s}$ is the number of chronon counts corresponding to one second, then $\Delta t$
must be divided by $n_\mathrm{s}$ to convert it into seconds. In conventional units,
equation (7) thus becomes 

\[
\Delta E\Delta t\sim mc^2/8n_\mathrm{s}.
\]
\vspace{-0.8cm}
\rightline{(8)}
 
\vspace{0.3cm} 
\noindent
Apart from an ambiguity relating to the mass $m$, this is precisely Heisenberg's 
uncertainty relation, if we identify Plank's constant $\hbar$ as

\[
\hbar=mc^2/8n_\mathrm{s}
\]
\vspace{-0.3cm}
or
\vspace{-0.3cm}
\[
n_\mathrm{s}=mc^2/8\hbar=2\pi c/8\lambda_\mathrm{c}
\]
\vspace{-0.8cm}
\rightline{(9)}
  
\vspace{0.3cm}
\noindent
where $\lambda_\mathrm{c}$ is the Compton wavelengh of the particle B. For an electron
the particle wavelength is $\lambda_\mathrm{c}=2.42\times 10^{-10}$ cm; equation (9 ) 
then gives a discrete time element of about $10^{-20}$ s. The corresponding length element
is just $c/n_\mathrm{s}\sim 8\lambda_\mathrm{c}/2\pi$, so that the space-time model under
discussion certainly fits in with the notion of continuity breakdown at the elementary
particle level. 

To generalise equation (7) to the case of a particle actually moving, in the sense 
that $k\not=1$, we note that the total energy $E$ is given by

\vspace{0.15cm}
\[
E=\half m[(n_\mathrm{r}/n_\mathrm{e})^{1/2}+(n_\mathrm{e}/n_\mathrm{r})^{1/2}].
\]

\noindent
We write

\[
n_\mathrm{r}=k^2n_\mathrm{e}+n
\] 

\noindent
where $n$ is the deviation of $n_\mathrm{r}$ from its mean value $k^2n_\mathrm{e}$,
and is of order $kn_\mathrm{e}^{1/2}$. Expanding $E$ in powers of $n$, we find 

\[
E=E_0+\half p_0(n/k^2n_\mathrm{e})+\textstyle{1 \over 8}(m/k-p_0)(n/k^2n_\mathrm{e})^2
\]
\[
~~~~~~~~~~~~~~~~~~~~~~~~~~~~~~~~~-\textstyle{1 \over 16}(2m/k-p_0)(n/k^2n_\mathrm{e})^3+...
\]
\vspace{-0.9cm}
\rightline{(10)}

\noindent
where

\[
E_0=\half (k+1/k)m~~~~~~\hbox{and}~~~~~~p_0=\half (k-1/k)m
\] 

\noindent
are the classical energy and momentum of the particle, i.e.~as revealed over long 
observation intervals.  On averaging over $n$, equation (10) gives

\vspace{0.15cm}
\[
\langle E\rangle=E_0+\textstyle{1 \over 8}(m/k-p_0)/k^2n_\mathrm{e}+...
\]
\vspace{-0.85cm}  
\rightline{(11)}

\noindent
and

\[
\Delta E^2=\langle E^2 \rangle-{\langle E\rangle}^2
\]
\vspace{-0.85cm}  
\rightline{(12)}
\[
~~~~~~=\left(\half p_0\right)^2/(k^2n_\mathrm{e})
+[\textstyle{1 \over 32}(m/k-p_0)^2-\textstyle{3 \over 16}p_0(2m/k-p_0)]/(k^2n_\mathrm{e})^2+...
\]

\vspace{0.3cm}
\noindent
The observation interval is $\Delta t=n_\mathrm{r}/n_\mathrm{s}\sim k^2n_\mathrm{e}/n_\mathrm{s}$; equations (9) and (12) together give

\vspace{0.15cm}
\[
\Delta E\Delta t
\]
\vspace{-0.85cm}  
\rightline{(13)}
\[
=[\left(4p_0/m\right)^2k^2n_\mathrm{e}+2(1/k-p_0/m)^2-12(p_0/m)(2/k-p_0/m)+...]^{1/2}\hbar
\]
\vspace{0.15cm}

\noindent
which is the full uncertainty relationship, accurate up to moderately relativistic
momenta.  Equation (13) implies

\vspace{0.15cm}
\[
\Delta E\Delta t~^>_\sim~[2(1/k-p_0/m)^2-12(p_0/m)(2/k-p_0/m)]^{1/2}\hbar.
\]
\vspace{0.15cm}
  
\noindent
The approximate equality holds if

\vspace{0.15cm}
\[
k^2n_\mathrm{e}<\textstyle{1 \over 8}[(1/k)(m/p_0)-1]^2
-\textstyle{3 \over 4}[(2/k)(m/p_0)-1]\sim \textstyle{1 \over 8}c^2/v^2
\]
\vspace{0.15cm}

\noindent
which is equivalent to

\vspace{0.15cm}
\[
\Delta t<{1 \over 2\pi}{c^2 \over v^2}{\lambda_c \over c}.
\]
\vspace{0.3cm}

A simple expression showing the deviation of B's motion from its classical path can 
be derived. We consider the expression (4) for $P_{n_\mathrm{e},n_\mathrm{r}}$ and use
Stirling's approximation: 

\vspace{0.15cm}
\[
n_\mathrm{r}!\sim (2\pi n_\mathrm{r})^{1/2}n_\mathrm{r}^{n_\mathrm{r}}\e^{-n_\mathrm{r}}.
\]
\vspace{0.15cm}

\noindent
For moderately large values of $n_\mathrm{r}$, equation (4) becomes

\vspace{0.15cm}
\[
P_{n_\mathrm{e},n_\mathrm{r}}\sim
(2\pi n_\mathrm{r})^{-1/2}\hbox{exp}[-\half(n_\mathrm{r}-k^2n_\mathrm{e})^2/k^2n_\mathrm{e}].
\]
\vspace{-0.825cm}
\rightline{(14)}

\vspace{0.45cm}
\noindent
We interpret this as follows: $n_\mathrm{e}$ is the interval between two light flashes
emitted by A, to make two determinations of B's position. For simplicity I shall assume
that the first is emitted as A and B pass (and hence returns immediately), and that A
chooses this event as his space-time origin. If the second position so determined has
coordinates $(t,x)$, we have

\[
n_\mathrm{e}=t-x 
\]
\[
n_\mathrm{r}=t+x
\] 
and
\[
n_\mathrm{r}-k^2n_\mathrm{e}=(k^2+1)(x-v_0t)
\]
\vspace{0.15cm}

\noindent
where $v_0=(k^2-1)/(k^2+1)$ is the classical velocity of B. Equation (14) thus becomes

\vspace{0.3cm}
\[
P_{t,x}=
[2\pi (t+x))]^{-1/2}\hbox{exp}[-\half(k^2+1)^2(x-v_0t)^2/k^2(t-x)].
\]
\vspace{0.3cm}
 
\noindent
This is the probability that after time $t$ the particle B is observed at position $x$;
we see that this position has a Gaussian spread about the classical path $x=v_0t$.

\section{Recapitulation}
 
Classical physics deals with systems having a continuous spectrum of states, and allows 
a system to be precisely located within this spectrum. In general quantum physics 
permits only a discrete spectrum, but allows the state of a system to spread over this 
spectrum according to a probability function. In this way classical and quantum 
physics can `correspond' in the appropriate limit. A similar contrast can be drawn 
between the classical picture of time and the one presented here; in this sense the 
theory discussed here is a quantum theory of time. 

The results presented in \S 6 relating to the motion of non-interacting particles are 
remarkably similar to quantum mechanical ones; it seems that any attempt to reformulate 
physics within this framework might start with classical physics, and might 
reproduce quantum theory, with hopefully some experimentally observable 
differences. The one unsatisfactory feature of the derivation of the uncertainty 
principle is the ambiguity relating to the mass $m$ of particle B. Equation (8) suggests a 
resolution of this ambiguity, which however involves taking seriously the chronon 
impact picture of time. If we regard the discrete elements of time as actual intervals 
between chronon impacts, we must specify a chronon register to be used in for 
example radar observations of particle motion. Clearly a macroscopic clock will 
experience many more such impacts in a given interval than say an electron, in direct 
proportion to the number of particles in the clock, i.e.~its rest mass. If we adopt the 
notion that when observing the motion of say electrons, then electrons are to be 
regarded as chronon impact registers, the resolution suggested by equation (8) is that 
the number of such impacts in a given interval is proportional to the rest mass of the 
particle in question, i.e.~that the rest mass is proportional to the chronon collision 
cross section. The ratio $m/n_\mathrm{s}$ is thus constant, and equation (8) is now an
unambiguous statement. If this idea is correct, it implies that space-time continuity breaks 
down at different levels for difierent species of elementary particle. An immediate 
consequence of this idea is that zero rest mass particles do not register chronons at all, 
which is quite acceptable as such particles follow null paths. 

There is of course a danger that the chronon gas would define an absolute rest 
frame, in the way that a classical gas of atoms or photons would. However, there is a 
fundamental difference here, which might prevent this. A particle moving slowly 
relative to the rest frame defined by say a classical relativistic gas would experience 
fewer atomic collisions per second than one at rest in this frame; by definition, the 
number of chronon impacts per unit time is the same for all particles of one species, 
whatever their relative motion. In any case, the chronon gas is not to be taken as 
filling space in the classical fashion, but as `being' space-time in some sense, along the 
lines of Penrose's idea of a `no space-time' model of space-time (Penrose 1967). 

In conclusion, we note that this discussion has been restricted to one spatial 
dimension, and to the one-dimensional motion of non-interacting particles. Clearly 
generalisations are required to produce a fully workable model. A generalisation to 
three spatial dimensions would involve a discussion of angles and directions, which 
might enable a theory of angular momentum to be developed. Gravitation, within the 
framework of classical general relativity, would probably be the easiest interaction to 
consider in this context, as there has been extensive work (Ehlers {\it et al} 1972;
Castagnino 1971) on the construction of curved space-time structure using light rays
and free particles.
\vspace{1.2cm}

\noindent
{\bf References}
\vspace{0.3cm}

{\obeylines\parindent=0pt
Bondi H 1965 {\it Brandeis Lectures, 1964} vol. 1 (Englewood Cliffs, NJ: Prentice-Hall)
\quad pp 386-399 
Castagnino M 1971 {\it J.~Math.~Phys.} {\bf 12} 2203 
Chew G F 1963 {\it Sci.~Prog.} {\bf 5l} October 529 
Ehlers J, Pirani F A E and Schild A 1972 {\it General Relativity: Volume Dedicated to
\quad J L Synge} (London: Oxford University Press)
Penrose R 1967 {\it Adams prize Essay: An analysis of the structure of space-time}
\quad (unpublished)
Schild A 1948 {\it Phys.~Rev.} {\bf 73} 414
}

\end{document}